# Measuring AI Innovation with Trademark Data

Carolina Castaldi[*] Fulvio Castellacci[†]
Andrea Fronzetti Colladon[‡] Ludovica Segneri[§]
Francesco Venturini[§¶]

July 9, 2026

**Abstract**

Researchers, managers and policy makers are exploring different approaches and data sources to map the development and the diffusion of Artificial Intelligence (AI). In this research note, we illustrate the opportunities offered by trademark data. We argue that AI trademarks can complement AI patents to capture different dimensions of AI innovation. AI trademarks can reveal the extent and ways in which companies exploit AI technologies to develop new goods and services. Importantly, trademark data offer a timely and globally available data source that covers all economic sectors. We present insights from using AI-trademarks in an empirical exploration of Italian firms. In our discussion, we reflect on how AI trademarks can be used at different levels of analysis to tackle emerging questions about the develop and diffusion of AI.

**Keywords**: AI; Trademarks; Patents; Innovation

**JEL codes:** O33; O34

[*]Utrecht University (NL). Correponding author: Vening Meineszgebouw A, Princetonlaan 8a, Kamer 6.66, 3584 CB Utrecht. Tel: +31 30 253 3205.

[†]TIK, University of Oslo (NO)

[‡]Roma Tre University (IT)

[§]University of Urbino Carlo Bo (IT)

[¶]NIESR and ESCoE (UK)

# 1. Introduction

Artificial intelligence (AI) is keeping researchers and policy makers busy. Hailed as the latest General Purpose Technology (GPT), AI is coming with a promise of profoundly transforming economies. However, the extent and nature of the actual transformation brought about by AI remains a matter of contention. A wealth of empirical studies is seeking to provide fresh evidence on the mechanisms at play, particularly when it comes to the implications of AI for a range of economic outcomes, from firm-level productivity to labor market dynamics. Understanding these implications requires measuring the many ways in which AI technologies are developed, adopted and diffuse across economic sectors. In this sense, empirical studies should be able to provide insight on patterns of innovation and adoption, ideally in ways that support replication and comparative analyzes. Unfortunately, despite the rapidly growing body of firm-level evidence, the results remain mixed, mainly due to the difficulty of accurately measuring firms' engagement in AI-related activities (Babina, 2026). Many studies have reported on general patterns of AI adoption,however, there is less evidence on the specific ways in which companies are developing novel AI-related technologies and products.

There are some fundamental measurement challenges surrounding AI innovation. In the context of AI and similarly to other GPTs, the distinction between the development, application, and adoption of the new technologies is often complex, as firms typically engage in these activities as part of a continuous process of innovation and learning. Some firms will innovate at the technological frontier, most other ones will engage in new-to-the-firm innovation to adapt AI technologies to their organizational needs (Calvino and Fontanelli, 2026) or will leverage AI technologies to develop new-to-the-market products. When it comes to AI innovation, most of the literature has used patent data to map AI innovations at different levels of data aggregation (Baruffaldi et al., 2020) and evaluate their economic impact (Marioni et al., 2024). However, patent activity only captures new-to-the-world innovation at the technological frontier and remains largely confined to large firms and specialized firms active in high-tech sectors.

In this research note we put forward the proposal of exploiting trademark data in a systematic and replicable way to measure product development activities part of AI innovation. Trademarks are filed by companies when introducing products to the market. We show how the so-called 'goods and service description' included in trademark records can be used to identify which trademarks refer to AI. AI trademarks reveal goods and services for which the company is using AI to develop a new market offering or upgrade an existing one. In this sense, they can reveal product innovations that are specifically enabled by AI (such as, for instance, a customized or enhanced product) or more generally represent applications of AI (e.g., analytical services based upon Large Language Models). Therefore, AI trademarks complement measures of AI innovation focused on technology development by focusing on product development and commercialization. Trademark data are particularly informative about technology follow-ups, incremental innovation, and the pervasive adoption of AI technologies (Jovanovic and Rousseau, 2005). Among the key features of trademarks is that they are used by companies of all sizes and ages and in all sectors of the economy, including services (Mendonça et al., 2004). Trademark data can be matched to firm-level data to study relationships between trademark activity and firm-level characteristics and performance. Moreover, trademark records offer both historical and timely data that are publicly available from national and international trademark offices.

Building on a first attempt to define and analyze AI trademarks by Nakazato and Squicciarini (2021) and Dernis et al. (2021), we develop a systematic approach that relies on refining and expanding the dictionary used to capture AI. We illustrate the usefulness of AI trademarks in an empirical analysis of Italian firms. Our analysis demonstrates how combining AI trademarks and AI patents allows drawing a richer picture of firms engaged in AI innovation. The analysis also sheds light on factors driving firm propensity to file AI trademark, compared to filing of AI patents. We find that firms filing AI trademarks are not the same firms that file AI patents, they are more numerous and they operate in different sectors with more representation of service sectors than those where AI patents concentrate. All in all, adding AI trademarks to the set of indicators used to measure firm-level AI innovation, helps to shed light on key mechanisms through which companies are exploring and exploiting the value of AI. Using trademarks is particularly valuable for studying AI because these data are both methodologically straightforward and widely accessible, making our paper a useful starting point for a broad range of future applications and extensions. Equally important, our empirical analysis accounts for the sparsity of AI filings, which represent rare events, as well as for the complex and largely unknown nature of the underlying innovation process, which is likely to be highly non-linear.

The rest of the paper is organized as follows. Section 2 reviews the literature on AI diffusion and innovation, and introduces trademark data, comparing them with other indicators. Section 3 explains our methodological approach to identifying AI trademarks. Section 4 details the setup of an empirical analysis of EUIPO trademarks filed by Italian firms. Section 5 reports the results of exploratory analyses that demonstrate the value added of using AI trademarks in firm-level analyses of AI innovation. Section 6 concludes by discussing opportunities and challenges for research, practice and policy alike.

## 2. Measuring AI innovation at the firm level

### 2.1. Firms and the diffusion of AI

The diffusion of AI is a major topic of research, as scholars try to make sense of the potential implications of AI technologies and measure their actual impact. In fact, the potential use and actual use of these technologies is often discussed as a key topic of research on its own (Babina, 2026). Similarly, the difference between the adoption of AI at the worker or task level and the firm-level adoption of AI flags crucial bottlenecks in the organizational diffusion of AI technologies. (Brynjolfsson et al., 2017). Nevertheless, a few stylized facts on which firms exploit AI are emerging.

First, although the advent of generative AI technologies has accelerated the diffusion of AI, its systematic and productive use remains relatively limited and concentrated in a small number of sectors (Ferrando et al., 2026). This indicates that the new GPT is still in its early phase of diffusion, with adoption and innovation concentrated in ICT manufacturing and service sectors (Calvino et al., 2022; Igna and Venturini, 2023). This evidence aligns with the longstanding notion that new technological paradigms take time to unfold and transform all economic sectors (David, 1990; Perez, 2010). Firms need to figure out the productive uses, often sector-specific, of the new technological options to really use AI in a transformative manner (Agrawal et al., 2019; Venturini, 2022).

Second, exploitation of AI varies considerably between firms and appears to be significantly correlated with firm size. As an intangible asset, companies must scale their investment to make it profitable (Bontadini et al., 2026). Large firms have more resources to adopt AI and experiment with it, so the link between AI and firm size might be mostly driven by self-selection. However, there is some evidence of a positive effect of AI on firm growth (McElheran et al., 2024; Brynjolfsson et al., 2021). At the same time, young and small firms also play a role (Calvino et al., 2022), particularly when it comes to the development of AI technology (Baruffaldi et al., 2020).

Third, not all AI users are alike (Calvino and Fontanelli, 2026). AI adoption includes a whole range of ways in which AI can be used within and by organizations, from scattered and incremental individual use by selected workers to improve tasks to more systematic organizational adoption leading to innovation. A useful distinction is the one between AI buyers and AI developers. The former simply buy AI technologies and tools developed elsewhere, the latter firms will instead develop new AI solutions and contribute to AI innovation. The expectation is that AI innovators should be the ones that experience the most evident economic benefits. Initial evidence on the link between AI innovation and firm-level productivity points to positive effects for firms at the frontier of AI technological development (Marioni et al., 2024). However, the evidence is still scattered.

Our interest here is exactly in AI innovation, understood as the outcome of developing new AI technologies or new products enabled by the application of AI (Babina, 2026). Most studies focused on AI innovation have exploited patent data and used different methods to identify AI patents as those linked to AI technologies (Giczy et al., 2022; Miric et al., 2023). However AI patents mostly capture technological innovations and these concentrate in patent-intensive technology sectors at the frontier of technology development. Patents are less suited to capture AI product development across the whole economy. Critically, the GPT nature of AI should appear exactly in the way these technologies allow a wide range of innovations throughout the economy (Jovanovic and Rousseau, 2005). These innovations might also be incremental, new-to-the-firm or new-to-the-market ones, but they are nonetheless crucial if AI is to really diffuse and have economic impact (Calvino et al., 2025). To capture AI innovations related to product development trademark data offers a number of opportunities, which we discuss next.

## 2.2. Introducing AI trademarks

Trademarks were originally proposed as novel data sources for innovation and industrial dynamics in seminal papers by Schmoch (2003) and Mendonça et al. (2004). The intuition was that the filing of a new trademark revealed the market introduction of a product, either good or service. As such, trademark filings can help track firm product strategies (Castaldi, 2020). At higher level of data aggregation, they can reveal market formation processes, national and regional specializations, and industrial dynamics (Mendonça et al., 2004; Castaldi and Mendonça, 2022).

A small but growing strand of research has built on that seminal intuition and provided original insights on how trademark data can be exploited, both as a complement to patent data and as an independent source of unique information on firm characteristics and performance (Schautschick and Greenhalgh, 2016; Castaldi, 2020; Willeke et al., 2026). There is also evidence that trademarks can be used as innovation indicators, helping to unveil commercialized innovation and non-technological innovation which are not covered by other data (Flikkema et al., 2014, 2019; Morales et al., 2024).

For the specific purpose of this paper, it is worth mentioning the few studies that already used trademarks to track the diffusion of new technologies. Jovanovic and Rousseau (2005) is an early exploration of new trademark filings as a proxy for product innovation activity connected to the emergence of new GPTs. In a similar vein, Gao and Hitt (2012) count new trademarks and find a positive correlation with the extent to which firms invest in ICT technologies, as these would increase opportunities for product variety. Babina et al. (2024) posit the same expectation in relation to AI: they find that one-standard-deviation increase in AI investment is associated with a 13% increase in new trademarks. The evidence is presented to support one of the mechanisms through which AI can boost economic performance of firms: by unleashing opportunities for product innovation.

These three studies align with the idea that trademarks can proxy product innovation and rely on counting all new trademarks filed after being exposed to the new technologies. The limitation of this approach is that overall trademark counts also include trademarks referring to innovation that is not related to AI. Therefore, identifying those trademarks specifically related to the application of AI will provide more precise information on which companies are exploiting AI to develop product innovation.

To address this issue, one can draw on pioneering work using the goods and service description (G&S description) included in trademark filings (Dernis et al., 2021). This textual information enables a more fine-grained characterization of trademarks and complements the broader information conveyed by the 45 Nice classes (i.e. those broad markets where the trademark applicant is seeking protection). It is fair to say that the G&S description is a so far underutilized element of trademark data. Most empirical studies using trademarks have focused on trademark counts or have classified them by market classes or type (see Schautschick and Greenhalgh, 2016 and Willeke et al., 2026 for an overview). A remarkable exception is von Graevenitz et al. (2022), which tracks keywords related to new technologies in the G&S description of trademark filings. Other relevant examples include works identifying specific types trademarks, including green or digital trademarks (Flikkema et al., 2025; Abbasiharofteh et al., 2026).

## 2.3. Comparing AI trademarks to other indicators of AI innovation

It is useful to compare trademark data to other data sources that have been used to derive indicators to measure AI innovation at the firm level. Table 1 summarizes the main differences between AI trademarks and the most used indicators: AI patents, survey-based indicators and webscraped indicators.

Patent-based indicators have been widely used to capture developments in AI technologies or analyze how AI might enable inventions in a range of technical fields (Cicerone et al., 2023; Xiao and Boschma, 2023). These studies exploited one of the many methods to classify AI patents (Ménière et al., 2017; Baruffaldi et al., 2020; Giczy et al., 2022). Studies based on AI patents typically show strong sectoral concentration of AI innovation, raising the question of whether this finding is driven by the patent-based measure or is corroborated by other indicators. Essentially, patent analyses provide evidence on AI technology development, while they can say little about AI product development. However, product development is a key element of AI innovation and the dimension of AI innovation that is closer to application and utilisation. In this sense, AI trademarks are an interesting alternative, given their much broader sectoral coverage.

At the same time, patent and trademark data share similar properties in terms of availability, comparability and possibility for matching to firm-level data. Intellectual property rights (IPR) data are easily available from IPR offices, often as historical records and in a timely fashion. In fact, market analysts tracking AI can rely on patents for technology intelligence and on trademarks for market intelligence (Lee and Lee, 2017; Castaldi, 2020). Clearly, the two also share similar limitations that accrue to IPR-based indicators, including missing innovation that remains secret or is simply not protected, and suffering from strategic use of IPRs (Castaldi et al., 2024).

Survey data have long been used to collect information on firms' innovation and technology adoption, capturing both their intensity and the underlying processes, including key drivers, barriers and effects (Castaldi, 2025). Surveys include administrative ones or ad-hoc surveys conducted in specific countries and periods. Survey data can offer detailed self-reported information on how firms integrate AI in their processes, also for young and small firms (McElheran et al., 2024). They have illuminated the distinction between AI buyers vs AI users (Calvino and Fontanelli, 2026) and those categories can help to understand differential impact on economic outcomes of AI adoption. Key weaknesses of survey-based indicators of AI innovation are their limited comparability over time and space, together with difficult matching to complementary firm-level data and weak timeliness.

An alternative approach relies on scraping corporate websites (Dahlke et al., 2024). This has the advantage of a much broader coverage than surveys, as webscraping can be done on the whole population of firms owning a corporate website. Therefore, this approach also allows covering small and medium sized firms, including those that do not fall in the category of AI startups (Baruffaldi et al., 2020). AI trademarks can reveal similar information as new product announcements on corporate websites, but their comparability is higher and costs are lower than webscraped AI innovation indicators.

Admittedly, one crucial challenge of all AI innovation indicators is the possible bias introduced by systematic "AI washing" practices, namely the standard practice of describing innovation as more valuable or novel than it really is. In practice all indicators discussed above suffer from this potential bias, but to different degrees. For AI trademarks, the fact that G&S descriptions are not written for customers might entail a lower risk of AI washing than the information on corporate websites. However, companies might use these descriptions strategically to signal their positioning to competitors or market analysts (Semadeni and Anderson, 2010).

The pioneering study on AI trademarks by Nakazato and Squicciarini (2021) is still the only attempt so far to define AI trademarks. The study first suggested identifying AI-related trademarks using AI-related keywords contained in G&S descriptions. Calvino et al. (2022) apply this approach to identify UK firms adopting AI, by combining trademark data with information extracted from patents and company websites. Building on these first attempts, our research note updates the methodology for identifying AI trademarks and examines the firm characteristics associated with AI trademarks as compared to AI patents.

Table 1: Firm-level indicators of AI innovation (own elaboration).

| **Indicators** | **IP indicators** | | **Other indicators** | |
|---|---|---|---|---|
| **Properties** | AI patents | AI trademarks | Survey-based indicators | Webscraped AI indicators |
| **Dimension of AI innovation captured** | AI technological invention | AI product innovation | Can cover technological and product innovation | Focus on AI product innovation, but also technological |
| **Coverage** | Patent-intensive sectors | All sectors | All sectors (if covered in survey design) | All sectors |
| | Large firms and startups | All firm sizes | All firm sizes (if covered in sample) | All firm sizes |
| **Comparability** | Historical data, allows comparisons over time | Historical data, allows comparisons over time | Limited temporal comparability unless explicit survey design | Limited temporal comparability |
| | Comparable across countries | Comparable across countries | Limited international comparability, unless explicit survey design | Comparable across countries |
| **Timeliness** | Timely, based on patent filings | Timely, based on trademark filings | Not timely, survey planning entails delay | Timely |
| **Availability** | Publicly available data from patent offices | Publicly available data from trademark offices | Data availability depends on survey but limited access | Publicly available company websites |
| **Costs** | Limited costs for indicators construction | Limited costs for indicators construction | Organizational costs to access data stored in secure environments | High costs of processing raw data |
| **Matching with firm-level data** | Matching already there for many countries | Matching can be done | Matching restricted because of anonymization | Matching can be done |

# 3. How to identify AI trademarks

## 3.1. Features of trademark records

Before delving into how one can identify AI trademarks, we briefly review the main features of trademark data as a means of Intellectual Property protection. Trademark records are published by national and international trademark offices after filing, and document the outcomes of the various stages in a trademark's lifecycle including abandonment, opposition, registration, and renewal (Graham et al., 2013). Trademark offices make these records available, also in bulk form, and update them regularly, often daily (Willeke et al., 2026). Hence, trademark data offers timely and detailed data including dynamic information about several characteristics of trademarks and their owners (Castaldi and Mendonça, 2022).

Figures 1 and 2 show two examples of trademark records from EUIPO (European Union Intellectual Property Office, the office managing EU-wide trademark filings). Example 1 covers CVskiller, an AI-based HR solution to evaluate skills included in CVs of job candidates. [1] Example 2 covers an Industry 4.0 solution to plastic packaging manufacturing.[2]

Upon filing, the applicant specifies the type of trademark that they wish to protect, i.e. any distinctive sign that helps consumers recognize the product and its origin. Figurative marks, like the examples in Figures 1 and 2, protect images, word marks protect names. Distinctiveness is only claimed in specific markets connected to 45 Nice classes, both goods and services. Example 1 covers Nice classes 9, 35, 42. Example 2 covers Nice classes 7, 16, 17, 39, 42. For each Nice class in which protection is sought, the applicant drafts a goods and service description that briefly details the product covered by the trademark. The exact wording can be updated upon feedback from the trademark office, to increase precision and ensure clarity of the legal protection claim. Some offices, including EUIPO, encourage using standard keywords in the G&S description, to speed up processing of trademark applications (Abbasiharofteh et al., 2026). Both examples in Figures 1 and 2 went through the EUIPO Fast track process. It should be noted that the G&S description varies in length and detail, as evidenced by our two examples. Its nature is also different from the technical text in patent records, where wording is crucial to prove novelty and industrial applicability.

[1] https://euipo.europa.eu/eSearch/#details/trademarks/018961112
[2] https://euipo.europa.eu/eSearch/#details/trademarks/018605323

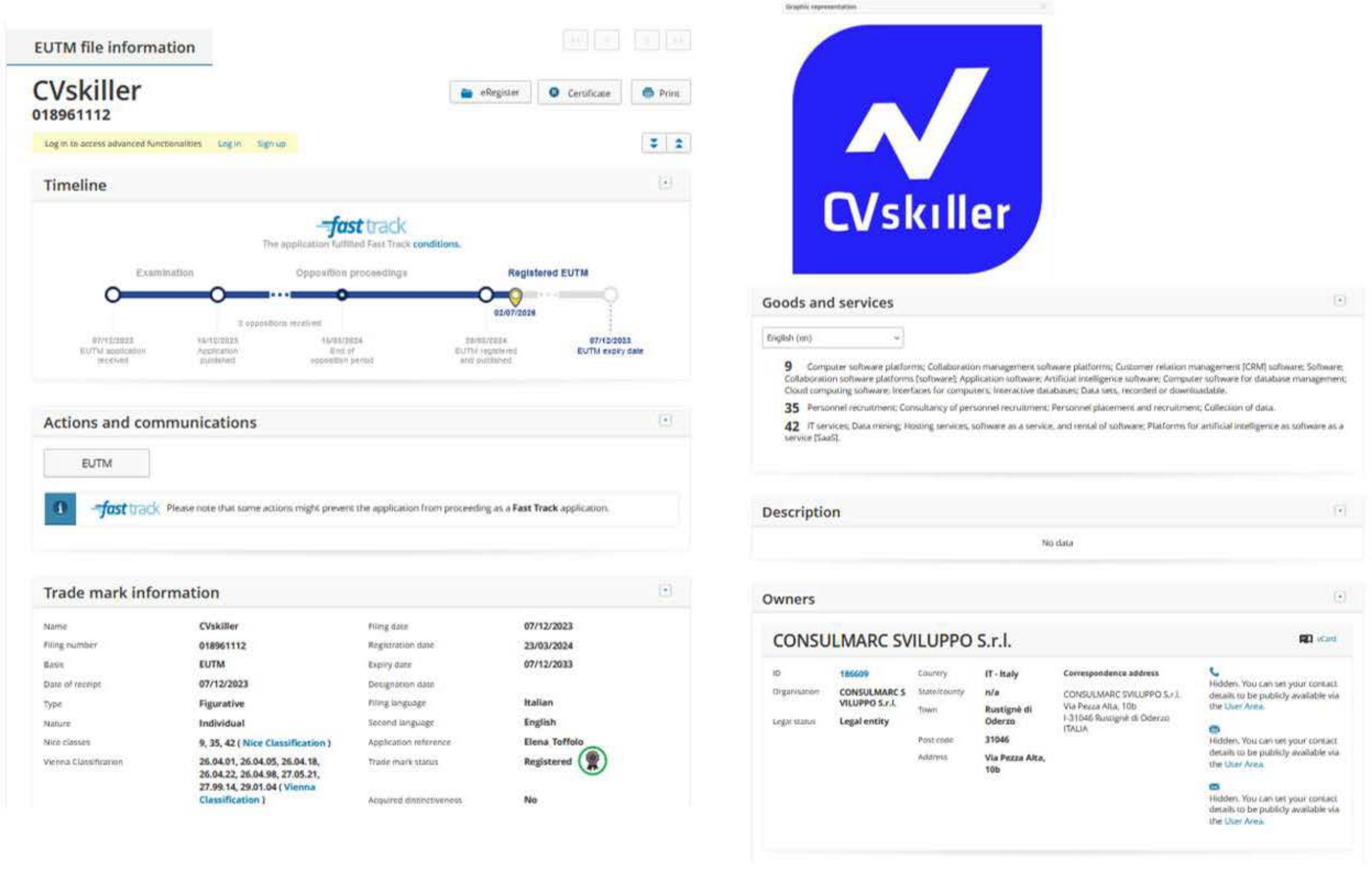

Figure 1: Example 1, selected elements from the screenshot from the eSearch Plus register at EUIPO, accessed 5 July 2026

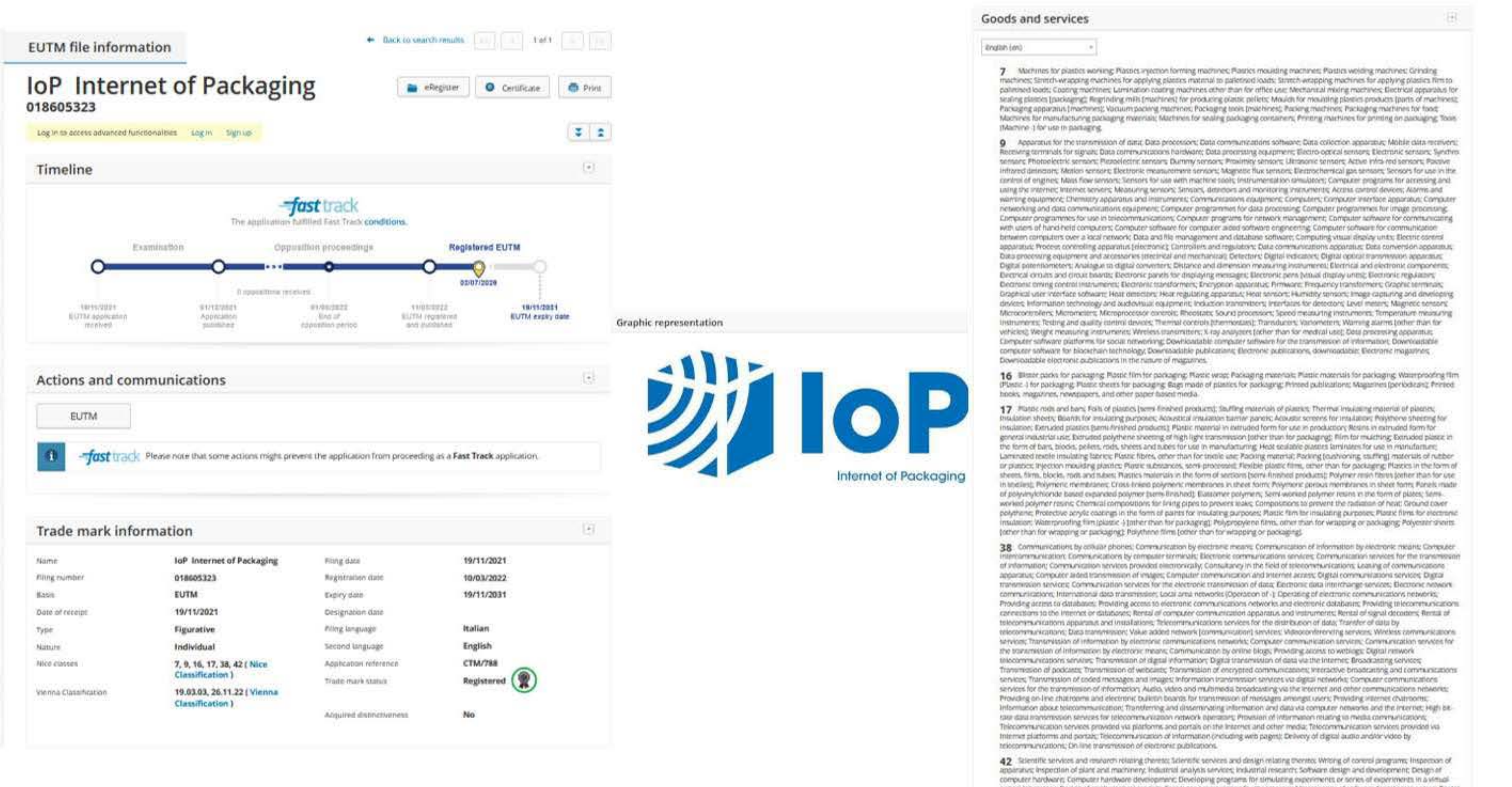

Figure 2: Example 1, selected elements from the screenshot from the eSearch Plus register at EUIPO, accessed 5 July 2026

### 3.2. Dictionary-based identification of AI trademarks

We follow Nakazato and Squicciarini (2021) and define trademarks as those that mention AI keywords in their G&S descriptions, relying on dictionary-based text matching. We opt for this strategy for its transparency, replicability and low computational costs. In the Appendix we show results from comparing its performance to more powerful but more computationally expensive approaches. An important clarification is that the most appropriate way to identify AI trademarks also depends on the research question guiding the analysis, and is bound to change over time as AI technologies continue to develop.

We update and extend the dictionary developed by Nakazato and Squicciarini (2021), which includes a list of AI-related keywords sourced from experts at the OECD, the Japanese patent office JPO, the UK intellectual property office IPO, and the World Intellectual Property Organization WIPO. As a first step, we remove terms that are prone to generate false positives, such as "k-means" or "logistic regression". These terms indicate statistical or analytical tools that were often associated with AI in the early stage of its development, but are now less likely to be considered AI products. Next, we expand the dictionary's coverage of more recent and application-oriented AI terminology, including terms related to generative AI, large language models, natural language processing, computer vision, speech technologies, robotics, autonomous driving, and AI explainability.

To do so, we generate a set of additional AI-related terms that are in use in technological and commercial contexts, as prompted by ChatGPT, Claude, and Perplexity. Next, we use the Lexicon Augmenter tool of the SBS BI app (Fronzetti Colladon and Vestrelli, 2025) to expand the vocabulary in two complementary directions. The first strategy queries WordNet to retrieve synonyms, hypernyms, and hyponyms for each input term (Miller, 1995). The second strategy uses pre-trained word embedding models to search for a pre-embedded vocabulary of WordNet terms and surface semantically similar words – capturing associations that are conceptually close but not connected through standard lexical relationships (e.g., "Generative AI" and "prompt" and "Large language Model" and "hallucination"). Finally, we collect information from online news to capture AI vocabulary in public and commercial discourse, including product categories, and application-specific terminology that may be particularly relevant in trademark descriptions. We apply topic modeling to the full text of online news of New York Times from 2023, extracted from the Event Registry platform. (Leban et al., 2014) We adopt a network-based approach in which words are connected by co-occurrence links and identify topics as communities within the resulting network (Fronzetti Colladon and Vestrelli, 2025).[3] The most relevant keywords are then extracted from this subset of articles. As the very final step, we clean the updated list to eliminate duplications and avoid false positives.

As a result of all the above steps, the final dictionary includes salient technical terms such as "neural network" and "artificial intelligence", as well as up-to-date product-related terms that cover the application of AI, such as "driverless". AI trademarks are those that mention at least one AI keyword in any of the Nice classes covered. Details of the validation procedure for the resulting dictionary are provided in the Appendix.

[3]After standard text pre-processing (Perkins, 2014), we removed low-weight links to retain only the most significant co-occurrences, and discarded nodes that are either isolated or excessively connected. We then performed community detection using the Louvain algorithm (Blondel et al., 2008), which accounts for edge weights when partitioning the network. We identified the top terms describing each topic using the importance metric proposed Fronzetti Colladon and Grippa (2020), which gives higher scores to words with strong internal connections to their cluster and weak connections to other clusters. "AI-relevant" articles where those in which topics accounted for at least 10% of the total word count.

# 4. An illustration: patterns of AI trademarks by Italian firms

We illustrate the information conveyed by trademark data and their potential to measure and analyze AI innovation by constructing and examining a novel dataset of Italian firms. We match trademark records with firm-level balance sheet data and also identify those firms that also file AI-patents. This allows us to compare our findings with existing patent-based evidence on AI innovation and to demonstrate the additional insights that trademark data can provide. After describing the database construction and presenting some descriptive results, we offer a machine learning analysis to study which Italian firms file AI trademarks and AI patents and what firm and sectoral patterns reveal about different aspects of AI innovation.

Italy provides an informative case study because it is one of Europe's largest economies, yet it is characterized by pronounced regional and sectoral disparities in innovation and economic dynamism, a relatively weak specialization in high-tech industries, and a large share of small and medium-sized enterprises operating in mature and traditional sectors. The Italian AI innovation landscape has already been studied using patent data (Lotti and Nobile, 2025), revealing the country's relatively weak position on the international frontier of AI technological development.

## 4.1. Database construction

We integrate three distinct data sources: (i) trademark filing data publicly available from EUIPO; (ii) patent filings at the European Patent Office extracted from Google patents, and (iii) company balance sheet information sourced from Moody's Orbis. The period covered is 2014–2023, since our last available year from Orbis is 2023.

We match trademark and patent data with the entire population of Italian firms included in Orbis using a fuzzy matching procedure based on applicant or company name, address, and postcode.[4] From the original Orbis sample of more than 397,000 companies, we identify 21,535 firm-year observations associated with trademark activity and 10,799 firms that filed at least one patent during the period under analysis. In the descriptive analysis reported in the next section we focus on these three groups of firms: those with trademarks only, those with patents only, and those with both trademarks and patents.

We classify trademarks as AI trademarks according to the approach outlined in the previous sections. The two examples reported in Figures 1 and 2 are both AI trademarks. Patents are classified as AI patents using a similar but more technology-oriented version of the dictionary used for trademark data. In what follows, we capture firms with new AI trademark filings in a given year with the firm-year dummy variable AI-TM, and those with AI patent applications with the dummy AI-PA.

We use balance-sheet information from Orbis to calculate several firm-level variables typically associated with innovation (see Gal, 2013; Andrews et al., 2019 for details). Value added and capital stocks are expressed at constant prices by deflating current-price balance sheet values using 2-digit NACE Rev. 2 industry deflators from EU KLEMS (Bontadini et al., 2023). Capital stock is constructed from fixed asset expenditure using the perpetual inventory method, applied to gross fixed investment and internal depreciation rates. Finally, we exploit the location and sectoral information of the IPR data to construct variables that account for potential geographical and sectoral externalities (both positive spillovers and business stealing effects) associated with AI trademark and patent activities. These are defined as the average count of AI-related IPR filings, patent or trademark, per firm within the same province (NUTS-3) or 2-digit industry (NACE Rev. 2). In both cases, the focal firm is excluded from the computation.

[4]We use harmonized information on the residence of trademark applicants included in EUIPO data. Instead, we retrieved information on the residence of patent applicants from the OECD PatReg database,

Figure 3: Number and share of firms owning AI trademarks or AI patents

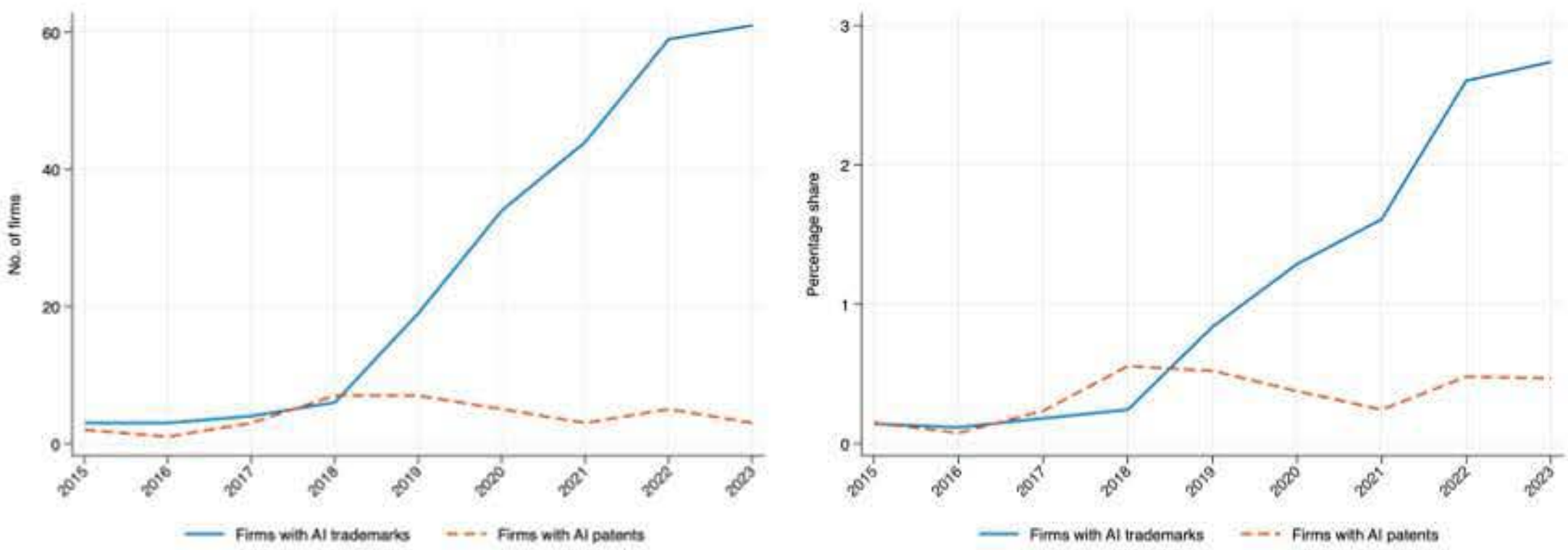

## 4.2. Descriptive analysis

AI innovation events are extremely rare in our sample of Italian firms: 238 AI-TM firm-years from 215 firms, and 36 AI-PA firm-years from 35 firms. Importantly, only 4 firms appear in both categories. Figure 3 illustrates the evolution, over the sample period, of the number of firms that applied for a new AI trademark each year compared with those that filed an AI-related patent. The figure also reports the corresponding shares of firms filing AI trademarks and AI patents relative to the total number of trademarking and patenting firms, respectively. In both absolute and relative terms, the numbers of AI-TM and AI-PA firms were small and broadly comparable during the first part of the sample period, remaining below ten firms per year. Since 2018 the number of AI-TM firms has increased substantially, reaching nearly 60 firms annually. This corresponds to approximately 3% of all trademarking firms in our sample. By contrast, the number of AI-PA firms has remained largely stable throughout the period.

Table 2 reports industry composition, (based on the NACE 2-digit classification) of the firms filing trademarks (or patents) and then the proportion of these firms that filed AI-trademarks (or patents). First, the overall number of firms filing trademarks is approximately twice as large as the number filing patents. This difference is particularly pronounced in the service sector, where trademarking activity substantially exceeds patenting activity, whereas the two figures are more comparable in manufacturing. More importantly, AI-TM firms are both more prevalent and more widely distributed across industries than AI-PA firms. Electricity and Computer Programs industries exhibit the highest incidence of companies with AI trademarks, while Computer Programs has the highest incidence of AI patenting firms.

Figure 4 shows the geographical distribution of AI-TM firms across NUTS3 regions, corresponding to Italian provinces. Firms filing AI trademarks are more widely dispersed throughout the country, showing a stronger presence in areas with relatively lower levels of patenting. In addition to the well-established innovation hubs around Rome, Milan, Turin, and Bologna, AI-TM firms are also found across several regions in the North-East and even in central and southern provinces. This broader spatial distribution sets the geography of AI trademarks apart from the more concentrated geography of AI patenting activity.

Finally, Table 3 reports the top three applicants for AI trademarks and AI patents. The table confirms the general patterns that the number of AI trademarks is substantially lower than the number of AI patents. However, among the leading AI trademark applicants, AI trademarks account for a relatively large share of their overall trademark portfolios. In contrast, for the leading AI patent applicants, AI patents constitute only a small fraction of their much larger patent portfolios. Notably, none of the top applicants appears in both the AI trademark and AI patent rankings. Among these firms, Techno, which is a company specializing in the design and manufacture of high-performance electrical connection solutions, is the only applicant that combines AI trademarking activity with broader patenting activity during the 2014-2023 period.

Table 2: Proportions of AI firms by sector

| | Firms with TM | % firms with AI TM | Firms with PA | % firms with AI PA |
|---|---|---|---|---|
| Mining and Quarrying | 13 | 0.00 | 14 | 0.00 |
| Food Products, Beverages and Tobacco | 2,194 | 0.05 | 136 | 0.00 |
| Textiles, Wearing Apparel, Leather and Related Products | 1,464 | 0.00 | 294 | 0.00 |
| Wood and Paper Products; Printing and Reproduction of Recorded Media | 361 | 0.28 | 171 | 1.17 |
| Coke and Refined Petroleum Products | 28 | 0.00 | 3 | 0.00 |
| Chemicals Products | 908 | 0.00 | 261 | 0.00 |
| Pharmaceutical Products | 304 | 0.33 | 179 | 0.00 |
| Rubber, Plastic and Non-metallic Mineral Products | 926 | 0.32 | 635 | 0.00 |
| Basic Metals and Metal Products | 1,149 | 0.17 | 1,159 | 0.17 |
| Computer, Electronics and Optical Products | 473 | 2.54 | 507 | 0.99 |
| Electrical Equipment | 433 | 1.85 | 377 | 0.00 |
| Machinery | 1,392 | 0.86 | 2,567 | 0.19 |
| Motor Vehicles and Other Transport Equipment | 280 | 0.36 | 294 | 0.34 |
| Furniture, Other Manufacturing and Repair | 1,024 | 0.68 | 579 | 0.17 |
| Electricity | 72 | 8.33 | 18 | 0.00 |
| Water Supply and Waste Management | 85 | 0.00 | 67 | 0.00 |
| Construction | 542 | 1.66 | 516 | 0.58 |
| Wholesale and Retail Trade of Motor Vehicles | 209 | 0.00 | 128 | 0.78 |
| Wholesale Trade | 3,771 | 0.61 | 859 | 0.35 |
| Retail Trade | 1,049 | 1.05 | 128 | 0.00 |
| Transportation and Storage | 251 | 1.20 | 58 | 0.00 |
| Postal and Courier Activities | 6 | 0.00 | 0 | 0.00 |
| Accommodation and Food Services | 642 | 0.16 | 100 | 0.00 |
| Finance | 123 | 0.81 | 45 | 0.00 |
| Publishing, Television and Broadcasting | 321 | 1.87 | 36 | 0.00 |
| Telecommunications | 35 | 0.00 | 9 | 0.00 |
| Computer Programming and Consultancy | 967 | 7.45 | 358 | 1.68 |
| Real Estate | 436 | 1.15 | 212 | 0.00 |
| Professional and Scientific Activities | 2,071 | 2.32 | 1,089 | 0.64 |
| TOTAL | 21,535 | 1.08 | 10,799 | 0.33 |

Figure 4: Geographical distribution of AI-TM and AI-PA firms by provinces (NUTS3)

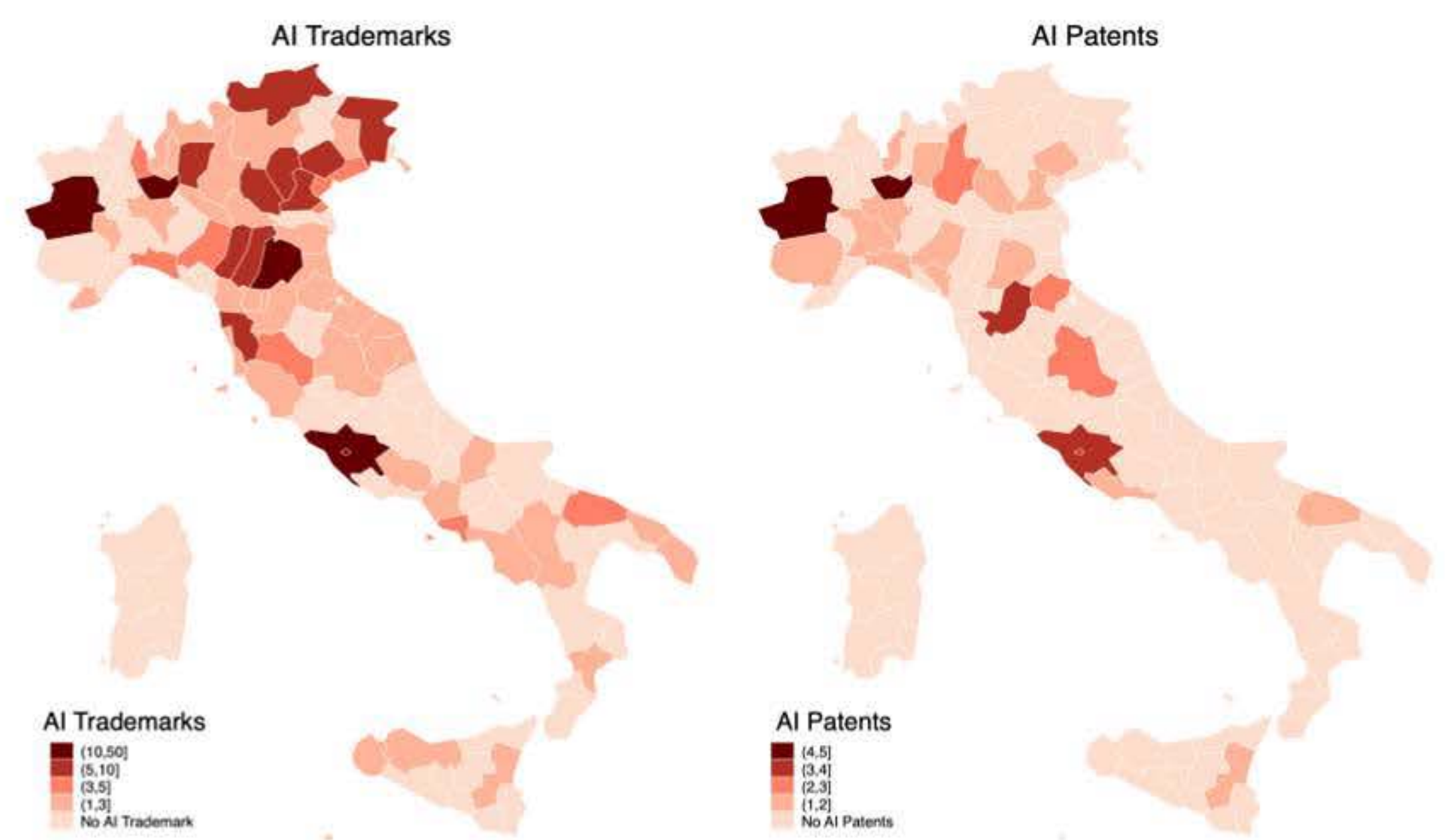

Table 3: Top AI applicants

| | Name | Province | Number of AI TM | Number of Total TM | Number of AI PA | Number of Total PA |
|---|---|---|---|---|---|---|
| | | | **Top 3 firms for AI Trademarks** | | | |
| 1) | Mundys. | Milano | 14 | 14 | 0 | 0 |
| 2) | Acea Energia | Roma | 13 | 22 | 0 | 0 |
| 3) | Techno | Como | 11 | 14 | 0 | 13 |
| | | | **Top 3 firms for AI Patents** | | | |
| 1) | Microservice | Torino | 0 | 0 | 115 | 5,749 |
| 2) | STMicroelectronics | Monza e Brianza | 0 | 0 | 105 | 5,165 |
| 3) | Metrologic Group Italia | Torino | 0 | 0 | 19 | 1,647 |

## 4.3. Machine learning analysis: comparing firms with AI trademarks and AI patents

The descriptive analysis above suggests that AI trademarks and AI patents follow distinct patterns. In this section, we demonstrate how a firm-level analysis of both can shed light on AI innovation. Using statistical machine learning methods, we examine which firms are more likely to file AI trademarks and AI patents, respectively, and identify the firm-level and system-level factors associated with each of these two distinct forms of AI-related innovation.

The reason for using a machine learning (ML) approach in this illustration is twofold. First, compared to standard parametric regressions, ML methods use a non-parametric approach that enables to take into account complex structures and possible non-linearities in the model. Since there is no yet well established knowledge about how to model the determinants of AI innovations, a ML approach is appropriate because it avoids imposing *ex-ante* parametric assumptions (e.g. linearity). Second, as noted in the descriptives above, AI innovations are still rare events, and the great majority of firms do not file patents or trademarks related to AI. To address the rare-event nature of AI innovations, our ML model assigns higher weight to innovating than non-innovating companies, an approach that is better suited to identify and classify this type of rare events than a standard linear or probit approach (King and Zeng, 2001 and He and Garcia, 2009).

The ML analysis proceeds in two stages. The first stage predicts which companies from the whole sample of Orbis firms files trademarks or patents related to AI innovations using two machine learning models: Lasso logistic regression and XGBoost gradient boosting. The two approaches are well-known methods of machine learning classification. The second stage uses the ML predictions to study how AI-TM firms and AI-PA firms differ. The Appendix provides further details on the methodology that we use to estimate and validate the ML model. In what follows, we focus on the results of this analysis and their main implications for the objective of this paper.

The variables that we use as predictors in these ML models include firm characteristics (productivity, size, age, labour intensity, cash flow), the firm's own IPR stock (cumulative TM and PA counts, capturing the firm experience with IPRs, overall and for AI) and spillovers from other firms' innovation activity at the provincial and sectoral level. All predictors are one-year lagged variables. The models also include year dummies, four macro-region dummies (with North-West as reference group), and sector dummies. Table 4 reports descriptive statistics of the variables used in this section.

Table 4: Variable definitions and descriptive statistics (whole sample)

| | **Variable** | **Mean** | **Min** | **Max** | **N (000s)** |
|---|---|---|---|---|---|
| *Outcome variables* | | | | | |
| | AI TM (firm-year) | 0.000 | 0 | 1 | 3,369 |
| | AI PAT (firm-year) | 0.000 | 0 | 1 | 3,369 |
| *Firm characteristics* | | | | | |
| | Labour productivity (log) | -6.34 | -37.4 | 73.9 | 3,194 |
| | Employment (log) | 1.64 | −29.11 | 10.90 | 3,328 |
| | Labour intensity (log) | -20.1 | -53.6 | -9.21 | 3,328 |
| | Cash flow ratio | 234.72 | 0.000 | 139.2M | 2,983 |
| | Firm age (log) | 2.65 | 0.000 | 5.14 | 3,328 |
| *IPR stock* | | | | | |
| | Own AI-TM stock (log) | 0.000 | 0.000 | 2.71 | 3,369 |
| | Own non-AI TM stock (log) | 0.026 | 0.000 | 4.69 | 3,369 |
| | Own AI-PA stock (log) | 0.000 | 0.000 | 4.66 | 3,369 |
| | Own non-AI PA stock (log) | 0.018 | 0.000 | 8.53 | 3,369 |
| *Provincial spillovers* | | | | | |
| | AI-TM activity in province | 0.077 | 0.000 | 1.59 | 3,335 |
| | Non-AI TM activity in province | 2.217 | 0.000 | 3.71 | 3,335 |
| | AI-PA activity in province | 0.161 | 0.000 | 2.35 | 3,335 |
| | Non-AI PA activity in province | 2.834 | 0.000 | 5.63 | 3,335 |
| *Sectoral spillovers* | | | | | |
| | AI-TM activity in sector | 0.073 | 0.000 | 2.82 | 3,335 |
| | Non-AI TM activity in sector | 2.040 | 0.000 | 6.16 | 3,335 |
| | AI-PA activity in sector | 0.173 | 0.000 | 2.84 | 3,335 |
| | Non-AI PA activity in sector | 2.966 | 0.000 | 6.64 | 3,335 |

To study the characteristics of firms that develop AI innovations, and how these characteristics differ between AI innovators (AI-TM vs AI-PA firms), we use SHAP (SHapley Additive exPlanations) values obtained from the XGBoost models, in which each firm's predicted innovation probability is decomposed into contributions from different predictor variables. These SHAP profiles are then used, together with simple descriptive statistics, to compare the characteristics of AI-TM and AI-PA firms.

Figure 5 shows the relative importance of the predictors in the ML models (longer bars indicate stronger prediction ability). As a first general finding, these models indicate that AI innovation is predictable. Results for AI-PA are weaker in the structural specification and should be interpreted with caution given the very small number of AI-PA firms. Nevertheless, several patterns emerge. First, labour productivity is the most important predictor for AI-TM, while for AI-PA the most powerful predictor is our proxy for sectoral spillovers in non-AI technologies, followed by labour intensity and productivity. This suggests that a firm's propensity to patent AI technologies depends more on being embedded in a sector characterized by higher patenting rates, reflecting more favorable technological opportunities and appropriability conditions, whereas its propensity to trademark AI-related activities depends more on the firm's own productivity. Second, sectoral knowledge spillovers display a clear pattern that highlights different determinants of AI patents and AI trademarks: the AI-TM model emphasizes the importance of sectoral spillovers associated with trademarks, while the AI-PA model relies most on patent spillovers. Third, firm age emerges as a meaningful predictor of AI-TM activity, but ranking lowest in the explanatory power, although this position is a bit lower for AI-PA.

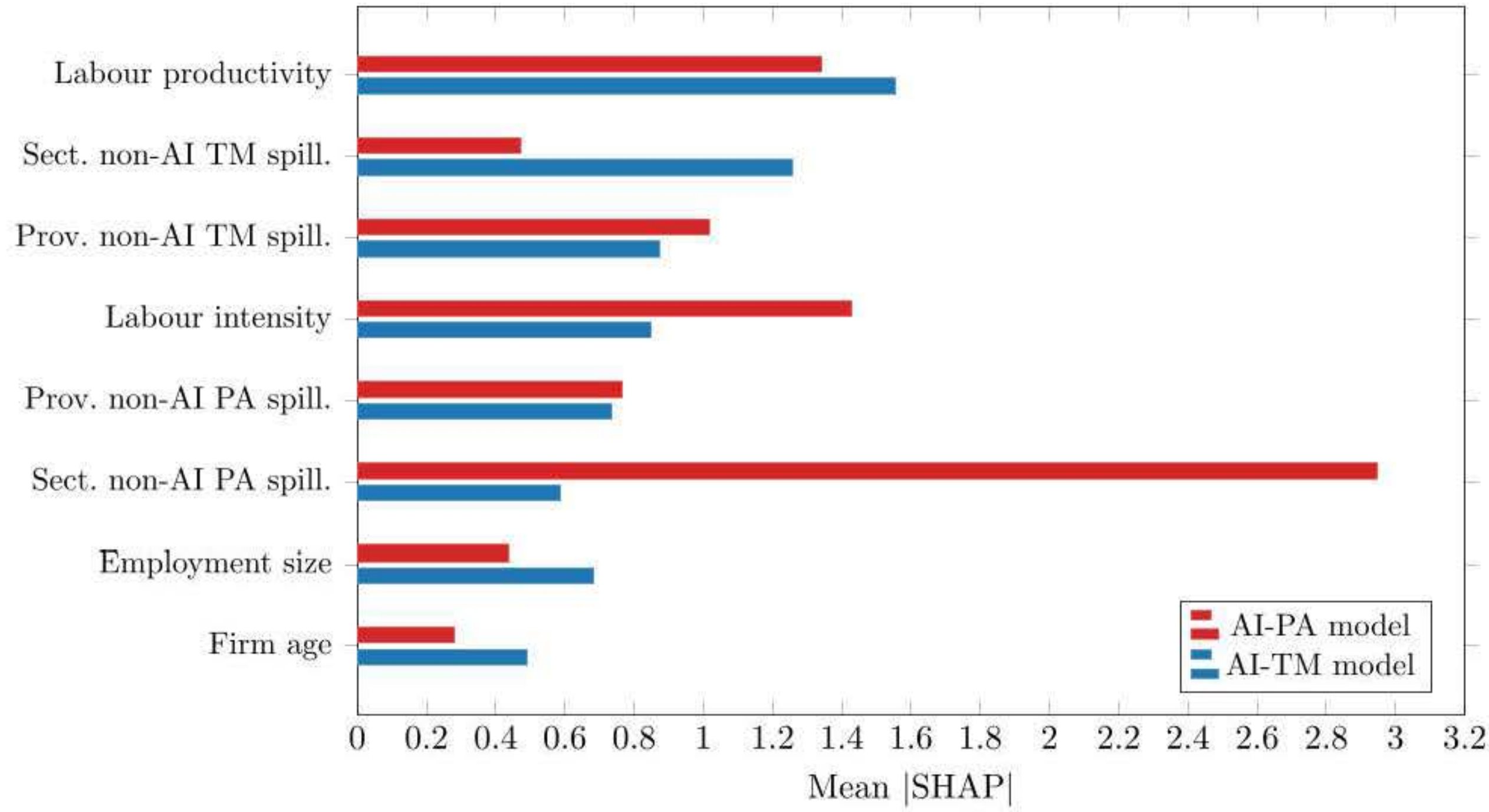

Figure 5: Structural SHAP importance: AI-TM vs AI-PA models. Each bar shows the mean absolute SHAP value for a given feature.

To complement the feature importance analysis, Table 5 reports the comparison between AI-PA and AI-TM firms. Three key findings stand out. First, the twp groups are nearly mutually exclusive in our data: considering the whole time span in our analysis, only four firms have filed both AI-TM and AI-PA. Hence, AI-TM and AI-PA flag two different innovation strategies. Second, AI-TM data reveals a much larger and broader population of AI innovators than what would be visible from patents alone: there are six to seven times more AI-TM firms than AI-PA firms in the Italian data. Moreover, these firms come from very different sectors. Third, AI-PA firms are more concentrated in advanced manufacturing (31% vs 18%), while TM firms are more concentrated in digital service sectors (35% vs 17%). These differences suggest that AI-PA activity tends to occur upstream in advanced manufacturing firms that develop novel methods, hardware, and processes; conversely, AI-TM activity is more frequent in firms that develop advanced digital services, as well as those that apply AI to bring their products and services to the market (see also Table 2). The sectoral pervasiveness in the diffusion of AI innovation as resulting from trademark data suggests that studies that rely on patent data alone tend to underestimate AI innovation in services, commercial activities and other medium and low-tech sectors, which are the sectors where much of the current firm-level AI innovation is actually taking place. Finally, AI-PA firms are on average slightly larger, older, and more productive than AI-TM firms, although the differences in firm characteristics between the two groups are not substantial.

Table 5: AI-PA firms vs AI-TM firms: key characteristics

| **Characteristic** | **PA firms (n=35)** | **TM firms (n=215)** |
|---|---|---|
| Mean size (log employment) | 3.01 | 2.81 |
| Mean labour productivity (log) | 4.22 | 4.17 |
| Mean firm age (log) | 2.92 | 2.74 |
| *Sectoral composition (% of firms):* | | |
| Advanced manufacturing | 31% | 18% |
| Digital services | 17% | 35% |
| Professional services | 20% | 19% |
| Trade | 11% | 15% |
| *Geography (% of firms):* | | |
| North-West | 49% | 36% |
| North-East | 17% | 37% |
| Centre | 29% | 21% |
| South / Islands | 6% | 6% |

**Note: Sectoral composition. Advanced Manufacturing:** Computer, Electronics and Optical Products; Electrical Equipment; Machinery; Motor Vehicles and Other Transport Equipment. **Digital services:** Publishing, Motion picture and video production, Television and Broadcasting; Telecommunications; Computer Programming and Consultancy. **Professional Services:** Professional and Scientific Activities: Legal and accounting, Management consultancy, Architecture and engineering, Scientific R&D, Advertising and market research, Other professional activities. **Trade:** Wholesale Trade; Retail Trade.

## 5. Discussion and conclusion

In this research note, we have argued that researchers and policymakers interested in measuring AI innovation can benefit from leveraging trademark data. The rationale is grounded in the classical literature on the diffusion of GPTs as breakthrough technologies (David, 1990; Jovanovic and Rousseau, 2005; Perez, 2010), which views the transformative potential of new technological paradigms as materializing through successive waves of innovation across the economy. While AI patents provide valuable information on technological inventions related to AI, they are less well suited to capture the development and commercialization of AI-enabled products and services across different sectors. In this respect, combining patent and trademark data offers a more comprehensive measurement and understanding of AI innovation, its development and applications.

Our illustration showed that AI innovation is still rare. This is in line with the idea that innovation is more than adoption. Innovation means that companies have been able to figure out how to use technologies productively in novel ways. This appears to be the case only for a very small percentage of firms, a reality check that resonates with results by others warning about the yet to be seen productivity effects of AI. As key input to this debate, we found that AI trademarks and AI patents are filed by similar types of firms, in terms of age and size, but these firms are structurally different in terms of sector. Interestingly, these are two distinct groups of innovating units: (almost) no firm combines the two IPRs suggesting that these activities are likely to be substitutes rather than complements at the firm level. As such, analyses that only rely on AI patents to measure AI innovation disregard a whole set of companies, concentrated in service sectors rather than advanced manufacturing ones.

These results, although exploratory, provide useful insights into the potential of trademark data to map AI innovation, alongside patent data. Still, our analysis is subject to two important caveats. In the first place, our analysis focused on Italy and was restricted to firms drawn from a selected database assembled to illustrate the potential use cases of trademarks. The observed distinction between firms with AI patents and those with AI trademarks may therefore reflect the absence of large AI firms that combine both technological invention and product commercialization. Older evidence presented in Dernis et al. (2021) suggests that AI trademarks and AI patents may coexist within the same firms in countries beyond those examined in the present study; however, such coexistence remains very rare. Such firms may be more prevalent in other national contexts, where large incumbents dominate the AI landscape and possess the capabilities to develop both AI technologies and AI-enabled products and services.

More general caveats go beyond the specificity of the Italian filings and concern general caveats of using IPR innovation indicators. Not all inventions and innovations are protected with IPRs. Therefore, a company that only files AI trademarks might still be developing AI technologies but not patent them, for different reasons ranging from the wish to avoid technological knowledge disclosure (Ouellette, 2012), speed to market (Leiponen and Byma, 2009), or other. Similarly, AI trademarks only cover a fraction of the many novel ways in which companies develop AI-related products. There is evidence on the share of total innovations covered by IPRs (Morales et al., 2024), but specific evidence on IPR strategies of AI firms is only emerging (Calvin and Leung, 2020; Rikap, 2024). IPR protection can also serve strategic purposes that clash with signaling innovation (Castaldi et al., 2024). Overall, more work is needed to study the IPR strategies around AI and to validate different approaches to define AI trademarks, also in comparison to complementary indicators of AI innovation. As we stressed in this note, our dictionary-based classification has a number of advantages (replicability, low computational cost, fit with current AI applications) but can updated and adapted in further and future research.

That said, the goal of this research note was to make the case for using trademark data, both by explaining how to identify AI trademarks in practice and analyzing their economic effects. The significant impact of firm-level and other contextual characteristics on trademarking behavior underscores the strength of the signal that AI trademark data can provide. We envision a whole set of interesting research directions that can be pursued by exploiting trademark data.

At the firm level, information on AI trademark activity can be linked to a broad range of economic variables, from productivity to labor market dynamics. Firms that register AI trademarks are more likely to develop AI-related products and increase their market share by responding to the increasing demand for AI applications. In this sense, a productivity premium is expected for these firms that goes through mechanisms of market differentiation. Demirev (2026) finds that increases in AI product innovation are positively associated with occupational wage premia and employment, though not uniformly across sectors. Similar ideas could be tested using AI trademark data on different samples of firms, across sectors and countries.

At a sectoral level, AI trademark activity could be related to different mechanisms of diffusion, to identify determinants of firm markups and changes to market structures. Greenhalgh and Rogers (2012) suggest that trademarks encourage imitation by revealing the market strategies of competitors in the same industry. This function can be particularly useful in the early phase of a GPT, when there is strong uncertainty about what to do with the new technologies. Of course, trademarks can also dampen competition in many ways, therefore an empirical question is whether AI trademark activity promotes dynamic competition even if it might increase static competition. A related mechanism is that, especially in digital markets, companies with successful AI product or innovation strategies can grow substantially and scale up globally.

From a geographical perspective, it would be interesting to deepen our understanding of the geography of AI innovation beyond the identification of AI technology strongholds (Xiao and Boschma, 2023). Regions might specialize in different dimensions of AI innovation, also those captured by AI trademarks and referring to new-to-the-firm innovation in non-technology intensive sectors. Lang et al. (2026) highlight processes of spatial decoupling in which countries might actually combine regional centers of specialization in different types of AI innovation, related to prior regional specializations. Therefore, these countries might build a multidimensional specialization in AI innovation, across different but complementary and connected regions, something that a simple regional analysis might not be able to detect.

At different levels of analysis, AI trademarks may provide novel ways to address critical questions about the impacts and costs of AI. These questions include: who is profiting from AI and who bears instead the costs, either economic, social or environmental ones? Importantly, trademark data can help map AI innovation across firms, sectors, and geographies and link it to a range of economic information in ways that also prompt normative questions about where the new technologies creates or destroys value (Castaldi, 2025). Such analyses appear particularly relevant for managers and policymakers alike. At a time when all economic actors struggle with the question of what AI technologies mean for them, it seems useful to have a realistic assessment of the economic opportunities and challenges around AI. For some firms or regions AI trademark data might allow revealing AI activities that that are already taking place but remain off the radar of other indicators. For others, they might help determine whether the costs involved in using AI are worthwhile or not. Overall, investment in AI might exacerbate existing disparities, between leaders and followers. Rodríguez-Pose and Bartalucci (2026) discuss these policy issues in the European context, warnig the AI investments might turn out to be orthogonal to cohesion policy goals. Minniti et al. (2025) study the effect of regional specialization in AI on factor income distribution, documenting reductions in the labour share of regional income, with consequences for regional income inequality. Clearly, the more these considerations can rely on indicators that capture different dimensions of AI adoption and innovation, the better for a meaningful assessment.

# Appendix

## Dictionary Validation

We validated our dictionary-based classification through a multi-step procedure combining large-scale automated checks with targeted human review.

As a first step, we ran a zero-shot classification using the GPT-4o-mini large language model on the full set of EUIPO trademarks with an application date between 1996 and 2024. The prompt instructed the model to classify a trademark as AI-related if its goods and services description referenced AI or machine learning technologies – covering terms such as artificial intelligence, neural networks, computer vision, natural language processing, speech and image recognition, autonomous systems, and AI-powered software – while explicitly excluding general software without AI features and products described merely as "smart" without evidence of AI capabilities. Since trademark descriptions frequently list multiple Nice classes, the prompt specified that a trademark should be classified as AI-related if any single class clearly mentioned AI technology.

The model agreed with our dictionary in 99% of cases, yielding a Cohen's kappa of 0.74 – a level of agreement generally interpreted as substantial. This provided initial reassurance about the reliability of the dictionary at scale, even when comparing groups of observations of very different sizes (in our case, AI trademarks being very few compared to non-AI trademarks).

To further test the classification, we randomly sampled 200 discordant observations, i.e., trademarks for which the dictionary and GPT-4o-mini produced opposite labels (100 positive and 100 negative). These cases constituted an adverse selection of the hardest-to-classify trademarks, and we subjected them to manual review by the authors. We additionally applied GPT-5.5 Extended Thinking to this same subset. Given the considerably higher computational cost of this model, applying it to the full corpus was not feasible within our budget constraints.

The results from this sample demonstrated that our dictionary performed well even in these borderline cases, achieving 81% accuracy and a Cohen's kappa of 0.62 relative to manual classification – a meaningful result given that the sample was explicitly selected as including difficult to classify cases. By contrast, GPT-4o-mini performed poorly (accuracy 19%, kappa –0.62), possibly also due to the length of some trademark records, which may saturate the model context window and degrade classification quality. GPT-5.5 Extended Thinking, on the other hand, aligned closely with human judgment even in ambiguous cases (accuracy 93%, kappa 0.84). Together, these findings suggest that our dictionary-based approach is sufficiently robust and offers a viable alternative to more expensive and less transparent approaches.

## Machine learning models used in Section 4: methodological details

For the analysis presented in Section 4 we trained two machine learning models, Lasso logistic regression and XGBoost gradient boosting, to predict AI trademark (AI-TM) and AI patent (AI-PA) activity, respectively. The two approaches are well-known methods of machine learning classification. Lasso logistic regression seeks to select the most relevant predictors by shrinking less important variables to zero; this is useful for identifying a few key variables that matter most in predicting a given outcome variable. XGBoost is a gradient boosting algorithm that builds a sequence of decision trees, where each tree corrects the errors of the previous ones. Differently from Lasso, it can capture non-linear relationships and complex interactions between variables.

We evaluated the predictive power of the models using two complementary validation approaches. A rolling temporal cross-validation consisted of five sequential folds: for each test year $t \in \{2017, ..., 2021\}$, the model was trained on all firm-years up to and including year $t-1$ and tested on year $t$. Performance metrics are reported for each fold and then averaged across folds. The final holdout consisted of training on all firm-years up to and including 2021 and testing on the two most recent years (2022 and 2023) jointly. In a nutshell, the rolling cross-validation gives a sense of how the model performs in repeated exercises across the years, while the holdout provides a single test on the most recent data.

Two complementary performance metrics were used to assess the predictive ability of these models, AUC-ROC and AUC-PR. AUC-ROC measures the model's ability to rank firms: if we pick one innovator and one non-innovator at random, how often does the model assign a higher score to the innovator? Values above 0.80 indicate good discrimination. Hence, this is an indicator of the model's ability to discriminate between AI innovators from non-innovators. AUC-PR summarizes the precision-recall tradeoff –that is more demanding when positive cases are very rare as in our dataset. The model's *precision* reflects what fraction of flagged firms are true innovators, while *recall* indicates what fraction of true innovators the model catches.

Table A1 shows that all ML models achieve very high AUC-ROC and perfect recall: every innovating firm in the holdout period is correctly identified. The Lasso is very selective; for the AI-TM model, it identifies only one non-zero coefficient: the firm's own AI-TM stock. This indicates that past AI trademarking is the dominant predictor in the model, much stronger than the other variables. The low precision of the indicator reflects the strong class imbalance: the 120 true positives are combined with 320 false positives among 732,000 test observations, yielding a false positive rate of just 0.04%. However, the results for AI patents (AI PA) in Table A1 should be interpreted with caution, since there are only 8 positive cases in the holdout set.

Moreover, in our sample only a small number of firms file AI trademarks more than once in different years during the time span, and only one firm files AI-patents more than once. Therefore, it is important to also test a model that removes the IPR stock (prior AI-TMs and prior AI-PAs), which in practice ensures that the regression sample also considers first-time AI innovators. As shown in Table A2, this model retains a high AUC-ROC, suggesting that firm characteristics and spillover variables carry an important predictive signal, even when prior accumulated IPR stock is excluded. The results for AI-PA are somewhat weaker and should be interpreted with caution given the very small number of AI-PA firms in the sample. AUC-PR collapses near zero because positive cases are extremely rare. Under such class imbalance, a model may still retain meaningful ranking ability, as indicated by AUC-ROC, while showing very weak precision-recall performance. The AUC-ROC values therefore suggest that these models still contain some ranking signal, although their ability to isolate rare positives is limited.

Table A1: Holdout performance (2022–2023). Baseline specification with IP stock

| **Target** | **Model** | **AUC-ROC** | **AUC-PR** | **F1** | **Prec.** | **Rec.** |
|---|---|---|---|---|---|---|
| AI-TM | Lasso | 0.9998 | 0.364 | 0.429 | 0.273 | 1.000 |
| AI-TM | XGBoost | 0.9998 | 0.285 | 0.429 | 0.273 | 1.000 |
| AI-PA | Lasso | 0.9998 | 0.026 | 0.074 | 0.038 | 1.000 |
| AI-PA | XGBoost | 0.9999 | 0.062 | 0.074 | 0.038 | 1.000 |

Table A2: Holdout performance. Specification without IP stock

| **Target** | **Model** | **AUC-ROC** | **AUC-PR** |
|---|---|---|---|
| AI-TM | Lasso | 0.863 | 0.002 |
| AI-TM | XGBoost | 0.807 | 0.002 |
| AI-PA | Lasso | 0.531 | 0.000 |
| AI-PA | XGBoost | 0.680 | 0.000 |

We also performed robustness tests to check the effects of different variables and model specifications. First, replacing provincial (NUTS-3) with regional (NUTS-2) spillovers produces almost identical results. AUC-ROC remains very high for both AI-TM and AI-PA. AUC-PR is slightly higher with regional spillovers, suggesting that the two geographic reference scales capture similar information. Hence, our findings are not sensitive to the choice of spillover aggregation level.

Second, we validated the model by holding out each macro-region in Italy in the cross-validation. The idea is to see whether the results hold when we train on some macro-regions and test on another macro-region. Table A3 reports these models' performance metrics. AUC-ROC is the same as before for every held-out macro-region, including when the entire North-West or North-East, which together contain the majority of AI innovators, are removed from the training data. For the South we find lower AUC-PR, reflecting the very small number of positive cases (15), but the discrimination ability of the model remains high. The spatial CV average closely matches the temporal holdout that was reported in the main exercise, confirming that the model learns firm-level and sectoral patterns rather than being affected by geographic clustering.

Third, we analyzed the model as a cross-section instead of a panel. Table A4 reports the results of this exercise. Collapsing the panel to a cross-sectional firm-level dataset again leads to high discrimination and higher AUC-PR for AI-TM, and lower for AI-PA. Averaging characteristics over time reduces noise, and the IPR stock becomes an even more important predictor at the firm level. Therefore, the cross-sectional results confirm that the panel findings are not an artifact of the temporal structure or trends in the data.

Table A3: Spatial cross-validation: AI-TM, XGBoost, baseline specification

| **Held-out region** | **Test TM+** | **AUC-ROC** | **AUC-PR** |
|---|---|---|---|
| North-West | 78 | 0.9999 | 0.401 |
| North-East | 83 | 0.9999 | 0.360 |
| Centre | 53 | 0.9999 | 0.289 |
| South | 15 | 0.9999 | 0.148 |
| Islands | 4 | 1.0000 | 0.351 |
| Average | | 0.9999 | 0.310 |
| *Temporal CV (reference)* | | *0.9998* | *0.285* |

Table A4: Cross-sectional ever-innovator analysis (5-fold × 5 repeats)

| **Target** | **AUC-ROC** | **AUC-PR** |
|---|---|---|
| AI-TM | $1.000 \pm 0.000$ | $0.865 \pm 0.068$ |
| AI-PA | $1.000 \pm 0.000$ | $0.385 \pm 0.137$ |